# Energy and equations of motion in a tentative theory of gravity with a privileged reference frame

M. ARMINJON (GRENOBLE)

**Abstract-** Based on a tentative interpretation of gravity as a pressure force, a scalar theory of gravity was previously investigated. It assumes gravitational contraction (dilation) of space (time) standards. In the static case, the same Newton law as in special relativity was expressed in terms of these distorted local standards, and was found to imply geodesic motion. Here, the formulation of motion is reexamined in the most general situation. A consistent Newton law can still be defined, which accounts for the time variation of the space metric, but it is not compatible with geodesic motion for a time-dependent field. The energy of a test particle is defined: it is constant in the static case. Starting from 'dust', a balance equation is then derived for the energy of matter. If the Newton law is assumed, the field equation of the theory allows to rewrite this as a true conservation equation, including the gravitational energy. The latter contains a Newtonian term, plus the square of the relative rate of the local velocity of gravitation waves (or that of light), the velocity being expressed in terms of absolute standards.

## 1. Introduction

AN ATTEMPT to deduce a consistent theory of gravity from the idea of a physically privileged reference frame or 'ether' was previously proposed [1-3]. This work is a further development which is likely to close the theory. It is well-known that the concept of ether has been abandoned at the beginning of this century. It is less widely known that, in the mean time, the objective situation in physics has made it reasonable and interesting to reexamine this concept. In such a matter, it is necessary for the writer to appeal to authorities. Thus:
(**i**) Due to the work of BUILDER [6-7], JANOSSY [11], PROKHOVNIK [21,23], it has now been proved in detail that special relativity (SR) is, after all, *fully* compatible with the assumption of the ether as envisaged by Lorentz and Poincaré, i.e. with the ether being an inertial frame in which Maxwell's equations can be written and relative to which all material objects undergo a 'true' Lorentz contraction. This has been emphasized by physicists as important as BELL [4], the author of theorems on 'hidden variables' in quantum mechanics. In a recent review on the latter subject, MERMIN [17] writes: « What Bell's Theorem did suggest to Bell was *the need to reexamine our understanding of Lorentz invariance* » (italics ours). It is worth to recall that several basic results of SR, including the Lorentz transformation and the Lorentz invariance of Maxwell's equations, were found by Lorentz and Poincaré, prior to Einstein, *within a theory based on ether*. As stated by BELL [4], the main differences between the approaches by Lorentz (and Poincaré) and by Einstein are that: (a) « Since it is experimentally impossible to say which of two uniformly moving systems is *really* at rest, Einstein declares the notions 'really resting' and 'really moving' as meaningless. » (b) « instead of inferring the experience of moving observers from known and conjectured laws of physics, Einstein starts from the *hypothesis* that the laws will look the same to all observers in uniform motion. » (italics are Bell's). It is now accepted that the Lorentz-Poincaré opposite philosophy leads to the same physical theory, namely SR.

(**ii**) It has become evident that quantum mechanics leads to attribute definite physical properties to the so-called 'vacuum', and that these properties do affect our measuring instruments as predicted by quantum mechanics. As SCIAMA [25] states: « For example, the electric and magnetic fields in the electromagnetic vacuum are fluctuating quantities. This leads to a kind of reintroduction of the ether, since some physical systems interacting with the vacuum can detect the existence of its fluctuations. However, this ether is Lorentz-invariant, so there is no contradiction with special relativity. »

(**iii**) Finally, relative velocities of astronomical objects are being measured with increasing accuracy. We can associate a privileged reference frame with the 'cosmic fluid', the velocity of which is the average velocity of matter. This reference frame appears clearly in the cosmological models because, in these simplified models, our universe is assumed homogeneous so that the average is already done. For example, the Robertson-Walker space-time metric of an expanding universe distinguishes a particular class of 'comoving observers'- that is, comoving with the expansion commonly assumed to explain, as a Doppler effect, the observed red-shift of the spectra emitted by distant galaxies. Their set makes a privileged reference frame, as emphasized by PROKHOVNIK [21-22].

Thus, we have a privileged reference frame in which also the 'empty' parts have physical properties. This may be called an 'ether' for short, even though 'physical vacuum' would be more appropriate. The use of the old name does not imply to forget modern physics: but that precisely today's physics allows to reconsider the assumption of an 'ether', provides some justification for reconsidering also the *possibility* that, after all, the allegable existence of this 'ether' *might manifest itself in certain laws of physics*. Since the privileged reference frame is kinematically defined from the average motion of matter at a very large scale, gravitation could be the range of such speculated manifestations. In other words, it seems allowable to investigate a theory of gravity with a privileged frame, one which would not necessarily be neutral in the theory: a non-covariant theory. Some important astronomical facts at the scale of galaxies, such as the famous rotation curves in the disc galaxies, have not yet received any undebatable interpretation [19,30]. Hence an explanation deriving from the effect of motion through an 'ether' might be considered.

However, Newtonian gravity (NG) gives an excellent description of astronomical motions at smaller scales, which is even refined by the corrections of general relativity (GR). Since neither NG nor GR does admit a preferred reference frame, it is undoubtedly a risky affair to attempt the construction of a theory of gravity with such a frame. A such attempt has yet been proposed [1-3]. The aim was to investigate whether the 'logic of absolute motion', which furthers our understanding of *special* relativity (SR) as compared with the usual 'space-time logic' [4,21], can be extended so as to build a sensible theory of *gravity*. The investigated theory is non-linear, like GR, and coincides with NG at the lowest approximation, also like GR. As an argument in favour of giving some attention to this theory, we mention the analytical study of the gravitational collapse in 'free fall' with spherical symmetry, i.e. the situation analytically studied within GR by OPPENHEIMER & SNYDER [18]. Since then, it has been shown, notably by PENROSE [20], that this situation contains all essential features of gravitational collapse for very massive objects in GR. According to the investigated theory, no singularity occurs during this collapse, neither for the metric nor even for the energy density, because the implosion would stop within finite time (for freely falling clocks and for remote clocks as well) and would be followed by an explosion [3]. On the other hand, this theory gives exactly the same predictions as GR for the motion of a test particle in the assumed static gravitation field of a spherical body, for in that case the Schwarzschild space-time metric of GR is obtained and the Newton law of the theory implies geodesic motion for any static field. The last result applies to mass particles [2] and to light-like particles as well [3], and provides a link between dynamics in NG and in GR, in the sense of MAZILU [16]. In this non-covariant theory, however, a gravitation field is static only if its source, the mass-energy density in the frame bound to 'ether', is time-independent. Whereas in NG and GR a uniform motion of the massive body can be eliminated by changing the reference frame, this is not the case here. Thus, in order to analyse either planetary motion in the solar system or stellar motion in a galaxy, one might have to reckon with velocities in the range 100-1000 km/s.

In the previous work, a Newton law has been defined only in the static case. Since it then implies geodesic motion, geodesic motion has been assumed in the general case. The purpose of this paper is to study in greater depth the formulation of motion in the theory, in connection with the problem of energy. We first recall (Sect. 2) the basic principles and equations of this theory, obtaining an alternative (equivalent) form of the field equations. For the equations of motion (Sect. 3), it is shown that in the general, non-static case, a consistent Newton law can still be defined, but is incompatible with the geodesic formulation of motion. Then the energy problem in the theory is analysed (Sect. 4). As is known, an exact energy can hardly be defined within GR, since the obtainment of a conservation equation for energy-momentum asks for a privileged class of reference frames (see e.g. STEPHANI [28]). After having recalled some aspects of the question of gravitational and total energy in NG and in GR, it is shown that a local energy balance equation, including the gravitational energy, is obtained with the studied theory if one assumes the (extended) Newton law of motion, but not under the geodesic formulation of motion.

## 2. **Basic assumptions and equations**
### 2.1 *Absolute space and physical vacuum*

Since astronomical observations and the success of Newtonian theory suggest that Galilean space-time is a very accurate description of our real world-theater, the notion of a physical vacuum has first to be conciled with *classical mechanics*. This, however, was not accomplished by ether theorists. For the ether must be the absolute space, hence rigid- otherwise, one would have *two* independent absolute spaces. But it also must offer no resistance against motion, and this in Newtonian mechanics can be true only of a perfect fluid. The proposed answer is that *it is the average motion of the hypothetical perfect fluid that defines the privileged inertial frame* [1-2], corresponding to the absolute space of Newtonian theory. Since any motion must be referred to some space (and time), it may seem that we are lead to a vicious circle. Thus we first postulate the absolute space or 3-D manifold *M*, which as a working assumption we may take to have Euclidean structure[1]. And we postulate the absolute time *t*, i.e. the world of events $M^4$ (time and position) is assumed to be the product **R**x*M* [2]. In Newtonian theory, the natural space metric $\mathbf{g}^0$ on *M* and the absolute time are assumed to be experimentally accessible, independently of motion and gravitation. In the studied non-linear theory, the coincidence between 'absolute' space (or time) metric and physical space (or time) measurements is found only in the first approximation. Then we may define the motion of a continuous medium, relative to *M*, by its velocity vector field $\mathbf{v}_0 = \mathbf{dx}/dt$. The particular metric structure we assume for *M* ensures that the integral of a vector field makes sense and is a vector (note that it is *not* the case for a general Riemannian space and thus for GR). Hence, the volume average of the velocity in any finite domain *Ω* occupied by the continuous medium is well defined, say $\overline{\mathbf{v}_0}^\Omega$. Now we may precise the assumption according to which the 'microscopic ether', that one which would be a perfect fluid, has in the average no motion relative to the absolute space *M* : we suppose that the velocity field of the micro-ether, $\mathbf{v}_{0e}$, is everywhere defined (it fills the space) and is such that $\overline{\mathbf{v}_{0e}}^\Omega \to 0$ as the size of the regular domain *Ω* (let us say it is a ball) infinitely increases, and this independently of the spatial position of *Ω*. We may call *M* the macro-ether.

Since the micro-ether must fill the space, any kind of matter, made of material particles, should be actually made of this universal fluid. Thus, material particles should be local, organized flows in ether, such as vortices which in a perfect fluid can be everlasting. We note that this concept is compatible with the lack of complete separability between particles, which is predicted by quantum mechanics and has received experimental confirmation. It would be an ambitious program, however, to attempt recovering micro-physics from this assumption of a '*constitutive ether*' (but not a hopeless program: cf. the hydrodynamic interpretation of Schrödinger's wave equation by MADELUNG [15]; more recently, ROMANI [24], WINTERBERG [31-32] and DMITRYIEV [9] obtained results in the same direction). Anyway, the gravitation theory needs almost only the macro-ether, which will be physically defined from the average motion of *matter*, i.e. the average motions of ether and matter are assumed identical. Recall that the notion of average used in this definition, is the asymptotic volume average. This means that no a priori bound needs to limit the size of local inhomogeneities, i.e. the size of astronomical structures.

---

[1] However, the whole theory can be written if *M* is more generally assumed to be equivalent (isometric) to either of Euclidean space $\mathbf{E}^3$, (hyper)sphere $\mathbf{S}^3$, or Lobatchevsky space $\mathbf{L}^3$, i.e. *M* is a space with constant curvature, be it nil, positive or negative: in either case, *M* is equipped with a natural metric $\mathbf{g}^0$ which admits a 6 parameter group of isometries.

[2] Formally speaking, one should start with $M^4$ and admit privileged space and time projections: $M^4 \to M$ and $M^4 \to \mathbf{R}$ making $M^4$ isomorphic to **R**x*M* in a canonical way. This is the 'Newton-Lorentz universe' considered by Soós [26].

### 2.2 *Scalar field equation and space-time metric*

A perfect fluid can exert only pressure forces. The gravitation force is tentatively interpreted as the pressure force or Archimedes' thrust, resulting from the gradient of the macroscopic pressure $p_e$ in the ether. This gives the following expression for the gravity acceleration **g** [1] :

$$(2.1) \qquad \mathbf{g} = -\frac{\mathrm{grad}\, p_e}{\rho_e},$$

where $\rho_e$ is the macroscopic density of the hypothetical fluid, assumed barotropic; the latter means that $\rho_e$ depends only on $p_e$, it is imposed to a true 'perfect' fluid which should be continuous at any scale, so that no temperature and no entropy can be defined [24]. The interpretation leading to Eq. (2.1) must be qualified as tentative in so far as it is based on the assumption that a micro-ether (or physical vacuum, or subquantum medium) does exist as a perfect fluid, whereas a consistent theory of microphysics following this line, still remains to build. In this interpretation, the gravitation force can be seen as a correction, i.e. it is what would remain when the microscopic fluctuations of ether pressure, which should be responsible for the other three forces, would have been accounted for. Although it is derived from this tentative interpretation, we take Eq. (2.1) firmly, but as a *phenomenological* equation in which $p_e$ plays the role devoted to the Newtonian potential $U$ in NG; it is thus substituted for
**g** = grad $U$. Just like for the latter, the macroscopic nature of Eq. (2.1) (the fact that the involved fields vary significantly over macroscopic distances only) does not prevent **g** from being felt at the scale of material particles.

The field equation for $p_e$, playing the role of Poisson's equation for $U$, is written as:

$$(2.2) \qquad \Delta p_e - \frac{1}{c^2} \frac{\partial^2 p_e}{\partial t_\mathbf{x}^2} = 4\pi G \rho\, \rho_e,$$

with $\Delta$ the Laplace operator, $\rho$ the density of matter (the density of the conserved mass in the first, Newtonian approximation; in the non-linear theory, $\rho$ is the mass-energy density in the frame E in which the points of $M$ have no motion[3]), $t_\mathbf{x}$ a *local time* to be precised hereafter, and $G$ Newton's gravitation constant. It is obtained basically from four requirements:

(i) Since NG propagates with infinite speed, it must correspond to the case where the fluid is incompressible, $\rho_e$ = Const., and the field equation must become equivalent to Poisson's equation as the compressibility evanesces.

(ii) In the compressible case, pressure waves must appear (except for static situations where $p_e$ does not depend on time), the velocity $c_e$ of which will be determined by the local ether compressibility, $K = 1/c_e^2 = d\rho_e/dp_e$.

(iii) Accounting for SR in the Lorentz-Poincaré interpretation thoroughly developed by PROKHOVNIK [21,23], the velocity of light, $c$, becomes a limiting speed: so should be also the velocity of the pressure waves (the 'sound' velocity) in the assumed constitutive ether, hence one must have $c_e = c$ everywhere (this implies that the barotropic relationship between $p_e$ and $\rho_e$ is *linear*, $p_e = c^2 \rho_e$).

(iv) In the theory, the equivalence principle arises naturally, as a correspondence between the absolute metric effects of motion and gravitation. As a consequence, *the space (time) standards are assumed to be contracted (dilated) in the gravitation field, in the ratio* $\beta = p_e/p_e^\infty$, where $p_e^\infty$ is the value of $p_e$ in regions which are remote from any matter and hence free from gravitation (see footnote 4, however); the space contraction occurs only in the direction of the gravity acceleration **g**. Hence, the 'physical' space metric **g** in the frame E becomes distorted, as compared with the metric natural $\mathbf{g}^0$, and *the* grad *operator in Eq. (2.1), as well as the Laplace operator in Eq. (2.2), are in terms of the curved physical metric* **g**:

$$(2.3) \qquad (\text{grad } \phi)^i = (\text{grad}_\mathbf{g} \phi)^i = g^{ij} \frac{\partial \phi}{\partial x^j}, \quad (g^{ij}) \equiv \mathbf{g}^{-1}$$

$$(2.4) \qquad \Delta \phi = \Delta_\mathbf{g} \phi = \text{div}_\mathbf{g} \text{grad}_\mathbf{g} \phi = \frac{1}{\sqrt{g}} \frac{\partial}{\partial x^i}\left(\sqrt{g}\, g^{ij} \frac{\partial \phi}{\partial x^j}\right), \quad g = \det(g_{ij}).$$

Moreover, a local time $t_\mathbf{x}$ appears at any fixed point **x** in $M$; thus if the point $\mathbf{x}_0$, also bound to $M$, is far enough from any matter so that no gravitation field is felt at $\mathbf{x}_0$:

---

[3] A (generally deformable) frame F can be defined as a time-dependent diffeomorphism $\psi_t$ of $M$ onto $M$, hence the frame E corresponds to the case where $\psi_t$ is the identity mapping at any time $t$.

(2.5) $$dt_\mathbf{x}/dt = dt_\mathbf{x}/dt_{\mathbf{x}_0} = \beta = p_e(t,\mathbf{x})/p_e(t,\mathbf{x}_0)\,,\qquad \beta = p_e/p_e^\infty < 1\,,$$

whence the definition of the derivation with respect to local time, appearing in Eq. (2.2):

(2.6) $$\frac{\partial \phi}{\partial t_\mathbf{x}} \equiv \frac{p_e^\infty}{p_e}\frac{\partial \phi}{\partial t} \equiv \frac{p_e(t,\mathbf{x}_0)}{p_e(t,\mathbf{x})}\frac{\partial \phi}{\partial t_{\mathbf{x}_0}}\,,$$

with $t_{\mathbf{x}_0} = t$ if is $\mathbf{x}_0$ is 'far enough', i.e. if $p_e(t,\mathbf{x}_0) \equiv p_e^\infty$. Note, however, that Eq. (2.6) assigns a unique value to any derivative with respect to local time, $\partial \phi / \partial t_\mathbf{x}$, even if one changes the reference point $\mathbf{x}_0$ in $M$, and this whether $\mathbf{x}_0$ is 'far enough' or not.

The assumed contraction of the physical space standards (and thus the dilation of measured distances), with respect to the natural metric on $M$, occurs in the direction of the gravity acceleration **g**. Hence the expression of **g**, the physical space metric in the frame E, is the simplest in an 'isopotential' coordinate system. This is a space-time coordinate system ($y^\alpha$) such that, at any given time $t$, $y^1$=Const. (in space) is equivalent to $p_e$=Const, and such that the natural metric $\mathbf{g}^0$ is diagonal: $(g^0_{ij})$ = diag $(a^0_1, a^0_2, a^0_3)$. In such a system, the gravitational space contraction implies that **g** also is diagonal, and writes:

(2.7) $\quad (g_{ij}) = $ diag $(\dfrac{1}{\beta^2}a^0_1, a^0_2, a^0_3) \equiv $ diag $(a_i)_{1 \le i \le 3}$ (isopotential system),$\quad \beta = p_e/p_e^\infty$.

(Greek indices will vary from 0 to 3 and Latine ones from 1 to 3). Note that, in the general case where $p_e$ depends on the time $t$, a such system will usually not be bound to the frame E, i.e. a point **x** in $M$ will in general have time-dependent coordinates $y^i(t,\mathbf{x})$. An important exception is the case where one has spherical symmetry around a fixed point $\mathbf{x}_0$ in $M$ (this either implies that $(M,\mathbf{g}^0)$ is Euclidean, or must be understood locally). Then, the time $t$ makes with the spherical coordinates $r$, $\theta$, $\phi$ an isopotential coordinate system bound to E.

The line element of the space-time metric **γ**, measuring the proper time $d\tau$ along an element of trajectory, follows straightforwardly from the combination of the slowing down of the mobile clock due to its absolute motion and to the gravitation field [2]. If $dl$ is the line element of the space metric **g**, i.e. if $dl$ is the elementary distance covered by the mobile, as measured with rods of the momentarily coincident observer bound to E, one has:

(2.8) $$ds^2 = \gamma_{\lambda\mu}\, dx^\lambda dx^\mu = c^2 d\tau^2 = \beta^2 (dx^0)^2 - dl^2, \quad x^0 = ct, \quad dl^2 = g_{ij}\, dx^i dx^j.$$

The validity of Eq. (2.8)$_3$ assumes that the coordinates ($x^\alpha$) are bound to the frame E. By Eq. (2.7), a diagonal expression is hence obtained also for $\boldsymbol{\gamma}$ in the particular case where an isopotential coordinate system bound to the frame E can be found. In that case:

(2.9) $$(\gamma_{\lambda\mu}) = \operatorname{diag}(\beta^2, -\frac{1}{\beta^2}a^0{}_1, -a^0{}_2, -a^0{}_3) \equiv \operatorname{diag}(b_\lambda)_{0 \le \lambda \le 3}.$$

If general coordinates ($z^\alpha$) are used, one writes instead of Eq. (2.8)$_3$:

(2.10) $$dl^2 = dt_\mathbf{x}^2\, g_{ij}\, v^i v^j = dt_\mathbf{x}^2\, v^2,$$

with $v^i$ the components of $\mathbf{v} = d\mathbf{x}/dt_\mathbf{x} = (1/\beta)\, d\mathbf{x}/dt$ (**v** is the velocity vector with respect to E and using the local time of the momentarily coincident clock of E) in the coordinates ($z^i$) (these being considered, at given time $t$, as a space coordinate system, thus $v^i = \dfrac{\partial z^i}{\partial x^j}\dfrac{dx^j}{dt_\mathbf{x}}$ if the ($x^\alpha$) are bound to E). Of course, $\boldsymbol{\gamma}$ is a space-time tensor and may be written in any space-time coordinates ($z^\alpha$). Also, a space metric **h** can be defined in the frame F bound to such coordinates, from the metric $\boldsymbol{\gamma}$: the components of the inverse matrix $\mathbf{h}^{-1}$ are $h^{ij} = -\gamma^{ij}$, with ($\gamma^{\lambda\mu}$) the inverse matrix of ($\gamma_{\lambda\mu}$). As shown by LANDAU & LIFCHITZ [12], the metric **h** is that one which gives the distances between neighbouring points bound to F (i.e. having constant space coordinates $z^i$), as evaluated from the interval of their local time, $d\tau$, for a to-and-fro light path: $dl' = (h_{ij}\, dz^i dz^j)^{1/2} = c\, d\tau/2$. Since SR still holds locally in the presence of gravitation, the result of PROKHOVNIK [21] is valid, according to which $dl'$ is indeed the physical distance which may be measured with rods of the frame F.

### 2.3 *Expression of the field equation in terms of natural metric and absolute time*

The expressions of the 'main' field equation (2.2) and the 'auxiliary' one (2.1) are in terms of the physical space metric and local (physical) time, in the frame bound to the absolute space or macro-ether M (it is recalled that Eq. (2.2) is only valid in this frame E). However, this physical space-time metric depends on the unknown, i.e. on the field $p_e$. In some cases, it is easier to handle Eq. (2.2) when it is expressed in terms of the natural metric $\mathbf{g}^0$ and absolute time $t$, as this turns out to be possible. Starting from Eq. (2.1) with $p_e = c^2\rho_e$, we use isopotential coordinates $(y^\alpha)$ and Eqs. (2.7) and (2.3), and obtain, since $p_e$ or $\beta$ depends only on $y^1$:

$$(2.11) \qquad \mathbf{g} = -c^2 \frac{\mathrm{grad}_{\mathbf{g}}\, p_e}{p_e} = -c^2 \frac{\mathrm{grad}_{\mathbf{g}}\, \beta}{\beta} = -\frac{c^2}{\beta}\frac{\beta^2}{a^0_{\;1}}\beta_{,1}\,\mathbf{e}_1 = -\frac{c^2}{2}\frac{1}{a^0_{\;1}}(\beta^2)_{,1}\,\mathbf{e}_1$$

where $(\mathbf{e}_i)_{1 \le i \le 3}$ is the natural basis associated with the coordinates $(y^i)$. Thus:

$$(2.12) \qquad \mathbf{g} = -\frac{c^2}{2}\,\mathrm{grad}_{\mathbf{g}^0} f = -\frac{c^2}{2}\,\mathrm{grad}_0 f, \qquad f \equiv \beta^2 = (\gamma_{00})_{\mathrm{E}}$$

(that $(\gamma_{00})_{\mathrm{E}} = \beta^2$ if one uses the absolute time coordinate $t$, is due to Eq. (2.8)$_1$). Turning to Eq. (2.2), we have $g \equiv \det(g_{ij}) = g^0/\beta^2$ from (2.7), thus we first obtain in the same way:

$$\Delta p_e \equiv \Delta_{\mathbf{g}} p_e = p_e^\infty \Delta_{\mathbf{g}} \beta = \frac{p_e^\infty}{\sqrt{g}}\left(\sqrt{g}\,\frac{\beta^2}{a^0_{\;1}}\beta_{,1}\right)_{,1} = p_e^\infty\,\frac{\beta}{\sqrt{g^0}}\left(\sqrt{g^0}\,\frac{\beta\,\beta_{,1}}{a^0_{\;1}}\right)_{,1}$$

(cf. Eq. (2.4)), which may be rewritten as:

$$(2.13) \qquad \Delta p_e = p_e^\infty\,\frac{\beta}{2}\,\Delta_0(\beta^2), \qquad \Delta_0 \equiv \Delta_{\mathbf{g}^0}\;.$$

Now, let us assume that our (model of) universe, in which the studied gravitation field and matter are embedded, is 'static at infinity', which is likely to be an extremely good approximation- except for cosmological problems. That is, assume that $p_e^\infty$ is independent of the time $t$ [4]. In that case, the term with time derivatives in Eq. (2.2) becomes:

---

[4] This assumption is in fact less restrictive than that of an 'insular matter distribution embedded in a Galilean space-time', which is commonly set in GR (e.g. FOCK [10], LANDAU & LIFCHITZ [12]). In particular, $p_e^\infty$ does not actually need to be reached somewhere, even asymptotically [3], and thus can be constant even if the matter distribution has unbounded support.

$$(2.14) \quad \frac{1}{c^2}\frac{\partial^2 p_e}{\partial t_x^2} = \frac{1}{c^2}\frac{1}{\beta}\frac{\partial}{\partial t}\left(\frac{1}{\beta}\frac{\partial p_e}{\partial t}\right) = \frac{p_e^\infty}{\beta}\left(\frac{\beta_{,0}}{\beta}\right)_{,0} = p_e^\infty \frac{\beta}{2f}\left(\frac{f_{,0}}{f}\right)_{,0}.$$

Combining Eqs. (2.13) and (2.14) and multiplying by $2/(\beta p_e^\infty)$, we rewrite Eq. (2.2) as:

$$(2.15) \quad \Delta_0 f - \frac{1}{f}\left(\frac{f_{,0}}{f}\right)_{,0} = \frac{8\pi G}{c^2}\rho.$$

Thus, the field equation reduces in the static case to the ordinary Poisson equation, which is *linear*, and indeed Eqs. (2.12) and (2.15) are equivalent, for time-independent $f$ (and $\rho$), to the Newtonian equations: exactly the Newtonian gravity acceleration **g** will be associated with any given density of mass-energy, $\rho(\mathbf{x})$, by Eqs. (2.12) and (2.15). However, in NG, only the 'invariable' (rest) mass is counted in $\rho$; moreover, it remains to precise the definition of $\rho$, and it turns out that $\rho$ must depend on the gravitation field, i.e. on $f$ itself [Sect. (4.2), point (iv)]. Anyhow, not the same motion will be predicted for test particles if **g** is known, since in the studied theory Newton's second law is expressed in terms of space and time measurements with clocks and rods of the local observer in E, which are affected by the gravitation field. In the static case with spherical symmetry, for example, Eq (2.15) (plus the requirement that **g** remains bounded as r $\to 0$, and the boundary condition $f=1$ at infinity), gives $f = 1 - 2\,Gm/(c^2 r)$ outside the body ($r$ being the radial Euclidean distance, and with

$$(2.16) \quad m = \int_M \rho\, dV^0.$$

where $dV^0$ is the volume element of the Euclidean metric). Then Eq. (2.9), with $a^0_1 = 1$, $a^0_2 = r^2$ and $a^0_3 = r^2 \sin^2\theta$ in spherical coordinates $r,\theta,\phi$, leads to Schwarzschild's exterior space-time metric. It is striking that, after natural account of SR and the equivalence principle, this theory (to which we have imposed to reduce to NG asymptotically as the 'ether compressibility' evanesces), gives exactly and simultaneously the Newtonian attraction field **g** and the Schwarzschild metric in the spherical static case [2]. The new result is that the Newtonian **g**-field is predicted for any static situation, whether 'spherical' or not. The form (2.15) of the field equation has important applications also in the non-static case, see Sect. 4.2 (energy balance).

## 3. The motion of a test particle: Newton law vs. space-time geodesics

To state Newton's second law demands to define the acceleration or rather (since SR must be taken into account) the time derivative of the momentum. Also because SR must hold at the local scale (and for sure with physical space-time metric), one has to use the physical, distorted space and time standards, i.e. the Riemannian space metric **g** and the local time $t_\mathbf{x}$.

### 3.1 *The case of a time-independent spatial metric* **g** *(static gravitation field)*

If on a manifold $M$ a Riemannian metric is given (thus a *fixed* metric **g**), one has a natural definition of the 'time' derivative of a vector $\mathbf{u}(t)$ attached to a 'trajectory' i.e. a differentiable mapping $t \mapsto X(t)$ from an open interval in **R** into $M$. The components $\eta^i$ of the derivative $D\mathbf{u}/Dt$ in a local coordinate system $(x^i)$ on $M$ are given by:

$$(3.1) \qquad \eta^i = \left(\frac{D\mathbf{u}}{Dt}\right)^i = \frac{d(u^i)}{dt} + \Gamma^i_{jk}\, u^j\, v_0^k \quad ,$$

with $\Gamma^i_{jk}$ the second-kind Christoffel symbols associated with metric **g** in coordinates $(x^i)$ and $v_0^k = dx^k/dt = (d\mathbf{x}/dt)^k$ the components of the 'velocity vector' (with respect to the 'time' $t$ which in (3.1) may be an arbitrary parameter, although in the rest of the paper $t$ denotes the absolute time). The definition (3.1) may be found in the literature, e.g. in BRILLOUIN [5] and in LICHNEROWICZ [13] where it is induced from that of a covariant derivative. However, a covariant derivative demands, strictly speaking, that **u** be a vector field (defined in an open domain around the considered point $X$), whereas here **u** is only defined along the trajectory, as a function of the parameter $t$ (and this, of course, is what is needed to use Eq. (3.1)). On requiring that the derivative $D\mathbf{u}/Dt$ cancel for any parallel vector and be a true derivative, i.e. obey the Leibniz rule, one is lead uniquely to characterize $D\mathbf{u}/Dt$ by the following intrinsic property:

$$(3.2) \qquad \text{for any vector } \mathbf{w} \text{ in the tangent space } TM_X \text{ at point } X \in M, \left(\frac{D\mathbf{u}}{Dt}\right).\mathbf{w} = \frac{d}{dt}(\mathbf{u}.\mathbf{w'}),$$

where point means scalar product **g** and **w'** is the parallel transport (using **g**) of vector **w** on the trajectory [2]; moreover, the components of the *vector D**u**/Dt* defined in this way, are indeed given by Eq. (3.1). Note that here $M$ is *also* equipped with the Euclidean metric $\mathbf{g}^0$ which would allow to speak of 'a vector **v** in $M$' (i.e. not specifically attached to a point $X \in M$), and in fact to identify points in $M$ and vectors (once an arbitrary origin point $O \in M$ has been selected) [5], whence our notation $\mathbf{x} \in M$ in previous sections. Thus the Newton law of SR has been extended to this theory of gravity, in the form:

$$(3.3) \qquad \mathbf{F} \equiv \mathbf{F}_0 + m(v)\,\mathbf{g} = \frac{D}{Dt_\mathbf{x}}(m(v)\,\mathbf{v}), \qquad \frac{D}{Dt_\mathbf{x}} \equiv \frac{1}{\beta}\frac{D}{Dt},$$

where $\mathbf{F}_0$ is the non-gravitational (e.g. electromagnetic) force, $\mathbf{v} = \dfrac{d\mathbf{x}}{dt_\mathbf{x}}$ is the velocity of the test particle with respect to $M$, $v = \mathbf{g}(\mathbf{v},\mathbf{v})^{1/2} \equiv (g_{ij}\, v^i\, v^j)^{1/2}$ its modulus and $m(v) = m(0)/(1 - v^2/c^2)^{1/2}$ is the inertial mass, which is thus identical to the passive gravitational mass, as is also true in NG. Although the latter identity is often referred to in GR (under the name of 'weak equivalence principle'), it does not make an exact sense in GR, because there is no Newton law there [27]. The Newton law (3.3) is also defined for *light-like* particles (photons, neutrinos?), in substituting the energy $e = h\nu$ (or rather $e/c^2$) for the inertial mass.

### 3.2 *Extension to the time-dependent situation*

In the general case, the metric **g** will yet depend on the time $t$ (i.e. its components $g_{ij}$ in coordinates bound to E will depend on $t$), hence it seems at first that Eq. (3.2), which is at the root of the definition of the time derivative of the momentum, used in Eq. (3.3), does not make sense any more: how can one define a parallel transport with respect to a metric that varies, i.e. with respect to a one-parameter family of metrics? Thus it seems at first that the Newton law (3.3) makes sense only for *static* gravitation fields, and so was it defined and used in the previous work [2-3]. However, it turns out that Eqs. (3.2)-(3.3) do still make sense, provided one 'freezes' the space metric $\mathbf{g}_{t_0}$ *of the time $t_0$ where the derivative is to be calculated*. In order to calculate the derivative $D\mathbf{P}/Dt_\mathbf{x}$ of the momentum $\mathbf{P}=m(v)\,\mathbf{v}$, one thus might use Eq. (3.1) (with $\mathbf{P}$ in the place of $\mathbf{u}$ and the local time $t_\mathbf{x}$ in the place of $t$) where the $\Gamma$ symbols refer to the metric $\mathbf{g}_{t_0}$. A natural definition of the parallel transport $\mathbf{w}'$ in Eq. (3.2) can indeed be given only if one considers the fixed metric $\mathbf{g}_{t_0}$. But this observation does not completely solve the question on the extension of Eq. (3.2) to the case with time-dependent metric $\mathbf{g}_t$: it remains to investigate whether the scalar product on the right of Eq. (3.2) should also refer to the fixed metric $\mathbf{g}_{t_0}$, or instead to the time-dependent metric $\mathbf{g}_t$. *If we take* $\mathbf{g}_{t_0}$, we define from (3.2) a derivative $D_0\mathbf{u}/Dt$ for which the Leibniz rule and Eq. (3.1) hold true, but clearly we are missing something, namely the time variation of the physical metric $\mathbf{g}_t$. *If we take* $\mathbf{g}_t$ and reexamine the derivation of Eq. (3.1) from Eq. (3.2) (Appendix 1 in [2]) we easily obtain, instead of (3.1):

$$(3.4) \qquad \left(\frac{D_1 \mathbf{u}}{Dt}\right)^i \equiv \frac{d(u^i)}{dt} + \Gamma^i_{jk}\, u^j\, v_0^{\,k} + g^{ij}\, \frac{\partial g_{jk}}{\partial t}\, u^k \equiv \left(\frac{D_0 \mathbf{u}}{Dt}\right)^i + \left(\mathbf{g}^{-1}\cdot\left(\frac{\partial \mathbf{g}}{\partial t}\cdot \mathbf{u}\right)\right)^i,$$

where, for a twice covariant second-order tensor $\mathbf{h}$ (here $\partial \mathbf{g}/\partial t$) and a vector $\mathbf{u}$, $\mathbf{h}.\mathbf{u}$ is the covector with components $h_{ij}\,u^j$. And, for a twice contravariant second-order tensor $\mathbf{k}$ (here $\mathbf{g}^{-1}$) and a covector $\mathbf{u}^*$, $\mathbf{k}.\mathbf{u}^*$ is the vector with components $k^{ij}\,u^*_j$. Thus, Eq. (3.4) also defines a vector $D_1\mathbf{u}/Dt$. Yet this time derivative, which accounts for the time variation of the metric, does not cancel for a vector $\mathbf{u}$ that is transported parallel to itself (necessarily with respect to the fixed metric $\mathbf{g}_{t_0}$) along the trajectory. The use of Eq. (3.2) for the definition is hence questionable, since Eq. (3.2) was obtained (in the case of a fixed metric) under the former requirement, plus the condition that the Leibniz rule has to be verified. If we provisionally forget the Leibniz rule, we can therefore define a one-parameter family of vector time-derivatives as well:

$$(3.5) \qquad \left(\frac{D_\lambda \mathbf{u}}{Dt}\right)^i \equiv \frac{d(u^i)}{dt} + \Gamma^i_{jk}\, u^j\, v_0^{\,k} + \lambda\, g^{ij}\, \frac{\partial g_{jk}}{\partial t}\, u^k \equiv \left(\frac{D_0 \mathbf{u}}{Dt} + \lambda\, \mathbf{g}^{-1}\cdot\left(\frac{\partial \mathbf{g}}{\partial t}\cdot \mathbf{u}\right)\right)^i$$

---

[5] In the case where $(M, \mathbf{g}^0)$ would be instead a space with constant curvature, one also might identify vectors in $TM_X$ and in $TM_O$ by the parallel transport along the geodesic of $\mathbf{g}^0$ joining $O$ to $X$. Contrary to the Euclidean case, this identification procedure allowing to speak of 'a vector **v** in $M$' would be dependent on the origin $O$, but in a way which should be harmless for the definition of a velocity field with *nil* macro-average (Sect. 2.1).

(that $D_0\mathbf{u}/Dt$ enters the definition (3.5) with the coefficient 1 is enforced, since we want to recover Eqs. (3.1) and (3.3) for a time-independent metric). Now let us come to Leibniz' rule.

We obtain from Eq. (3.5) and the fact that, by construction, $D_0\mathbf{u}/Dt$ obeys the Leibniz rule with the fixed metric $\mathbf{g}_{t_0}$:

$$\mathbf{u}.\frac{D_\lambda \mathbf{v}}{Dt} + \mathbf{v}.\frac{D_\lambda \mathbf{u}}{Dt} = \mathbf{u}.\frac{D_0 \mathbf{v}}{Dt} + \mathbf{v}.\frac{D_0 \mathbf{u}}{Dt} + 2\lambda \frac{\partial \mathbf{g}}{\partial t}(\mathbf{u},\mathbf{v})$$

(3.6)
$$= \frac{d}{dt}\left[\mathbf{g}_{t_0}(\mathbf{u}(t),\mathbf{v}(t))\right] + 2\lambda \frac{\partial \mathbf{g}}{\partial t}(\mathbf{u}(t_0),\mathbf{v}(t_0)).$$

But, by direct calculation, we also have:

$$\frac{d}{dt}\left[\mathbf{g}_t(\mathbf{u}(t),\mathbf{v}(t))\right] = \frac{d}{dt}\left[g_{ij}(t,\mathbf{x}(t))\,u^i(t)\,v^j(t)\right] = \frac{d}{dt}\left[g_{ij}(t_0,\mathbf{x}(t))\,u^i(t)\,v^j(t)\right] + \frac{\partial g_{ij}}{\partial t}u^i(t_0)v^j(t_0),$$

(3.7) $$\frac{d}{dt}\left[\mathbf{g}_t(\mathbf{u}(t),\mathbf{v}(t))\right] = \frac{d}{dt}\left[\mathbf{g}_{t_0}(\mathbf{u}(t),\mathbf{v}(t))\right] + \frac{\partial \mathbf{g}}{\partial t}(\mathbf{u}(t_0),\mathbf{v}(t_0)).$$

By comparing (3.6) and (3.7), we find that *Leibniz' rule holds (with the variable metric $\mathbf{g}_t$) if and only if $\lambda=1/2$ in Eq. (3.5)*. Nevertheless, with any value of $\lambda$, we may associate a particular Newton law, in the following way:

(3.8) $$\mathbf{F} \equiv \mathbf{F}_0 + m(v)\,\mathbf{g} = \frac{D_\lambda}{Dt_\mathbf{x}}(m(v)\,\mathbf{v}), \quad \frac{D_\lambda}{Dt_\mathbf{x}} \equiv \frac{1}{\beta}\frac{D_\lambda}{Dt}, \quad \mathbf{v} \equiv \frac{d\mathbf{x}}{dt_\mathbf{x}} \equiv \frac{1}{\beta(t,\mathbf{x})}\frac{d\mathbf{x}}{dt}.$$

Now is $\lambda=1/2$ really the correct value for the parameter $\lambda$? Another possible criterion for the choice could be the compatibility of Eq. (3.8) with the formulation of motion in GR. It has been proved that, *in the static case*, the Newton law (3.3) *implies* that any free test particle follow a geodesic line of the metric $\boldsymbol{\gamma}$, and this for a mass particle [2] as well as for a light-like one [3]. It is not difficult (though a bit tedious) to follow the proof for a mass particle in the static case, the calculation method also applying to the case where one still has isopotential coordinates bound to E. One then finds that, already in that particular case, any time dependence of the field $p_e$ makes new terms appear in the geodesic equation for the values 0 and 1 of index $\alpha$, corresponding to new non-zero Christoffel symbols of metric $\boldsymbol{\gamma}$ in isopotential coordinates; and these new terms imply that, for $\alpha=0$ and $\alpha=1$, this equation $G^\alpha = 0$ is verified if and only if $\lambda=1$. The $\alpha=0$ term will be examined indirectly in Sect. 4.2 (in connection with the energy balance). In summary, we presently cannot state for sure that the Newton law (3.8) with $\lambda=1$ implies geodesic motion in the most general case, but we already know that *the law (3.8) with any $\lambda\neq 1$, in particular with $\lambda=1/2$ (as is needed to have Leibniz' rule) is incompatible with geodesic motion*. Here is thus a crucial bifurcation in the studied ether theory: *one has to decide between Newton law (with Leibniz rule, thus $\lambda=1/2$) and space-time geodesics*. In our previous work, the second choice had been made. This choice is consistent with the commonly accepted 'logic of the space-time', but this was not the main reason for the choice: simply, no other possibility had been seen! From a rather philosophical point of view, the first choice would be in fact more consistent with the 'logic of absolute motion' which is studied here (cf. PROKHOVNIK [21] for a detailed discussion of the two different 'logics' in the frame of SR). The agreement with observation might be the true judge between the two possibilities, both of which are indeed compatible with the kernel of the studied theory (summarized in Sects. 2.1 and 2.2) from a purely physical viewpoint. A strong physical argument for the choice 'Newton law' (with $\lambda=1/2$) and against 'Einstein geodesics' within the studied theory will, however, appear in Sect. 4.2.

## 4. The energy problem in the studied theory of gravitation

### 4.1 *Some remarks on energy and conservation laws in Newtonian theory and in general relativity*

(**i**) Is the concept of energy a relative one?

The concept of energy is among the most important one in today's physics, especially in classical physics (classical mechanics of mass points and continuous media, thermodynamics, classical electromagnetism) and in microphysics as well, but the notable exception is the gravitation theory (GR): there, this concept can have only an approximate status since it is not a covariant concept. Already in NG, the (kinetic plus potential) energy of a mass point, $e = (v^2/2 - U)m$, is not even a Galilean invariant. Indeed, $v$ is obviously not invariant, whereas the potential *is* a Galilean invariant, since both Poisson's equation and the merely spatial boundary conditions, e.g. $rU$ and $r^2$.grad $U$ bounded at infinity, are so (it is recalled that strict NG imposes a spatially bounded distribution of mass). That $e$ is changed by a constant when changing the inertial frame makes, of course, no problem in NG since only the variations of $e$ are relevant there. But since NG demands a bounded mass distribution, the energy in the frame of the mass-center *could* be preferred on theoretical grounds, and is indeed preferred in actual analyses: in classical mechanics, be it celestial or terrestrial, one considers an assumed isolated system and one refers the velocity, and hence the kinetic energy, to the mass center. Strictly speaking, one should consider an (approximately) isolated subsystem of the assumed bounded universe and, for elements of the subsystem, evaluate their velocity in the global mass-center frame.

But this would differ by a *definite* constant (the velocity of the mass center of the subsystem) from the velocity in the frame bound to the mass center of the subsystem. Thus in NG (and in the same way in a good part of classical physics), an absolute concept of energy is allowable, probably favourable indeed (cf. the definition of temperature in statistical thermodynamics, as a mean kinetic energy). Apart from this, it seems that, in non-relativistic quantum mechanics, one would prefer to avoid discussing the effect, on the energy levels, of changing the reference frame. It may be that in quantum mechanics also, an absolute concept of energy would be favourable.

(**ii**) Energy and conservation laws in Newtonian gravity

In addition to this (debatable) absolute character, a still more important (and undebatable) aspect of the energy concept is that, in classical physics, it gives rise to local *balance equations* for continuous media, which lead to global *conservation laws*. We take the example of NG, which is relevant here, and we assume elastic behaviour for simplicity (this includes the case of a perfect barotropic fluid). Thus one has, in NG, the following definition and conservation equation for the volume density $w$ of the total energy, i.e. the energy of matter (including its potential energy in the gravitation field), $w_m$, plus the energy of the gravitation field itself, $w_g$:

$$(4.1) \qquad w = w_m + w_g, \quad w_m = \rho(v^2/2 + \Pi - U), \quad w_g = \frac{\mathbf{g}^2}{8\pi G},$$

$$(4.2) \qquad \frac{\partial w}{\partial t} + \operatorname{div}\left(w_m \mathbf{v} - \boldsymbol{\sigma}.\mathbf{v} - \frac{\partial U}{\partial t}\frac{\mathbf{g}}{4\pi G}\right) = 0,$$

with $\boldsymbol{\sigma}$ the stress tensor and $\Pi$ the mass density of elastic energy; this is derived from the more usual energy balance (in which the energy $w_g$ does not appear and a source term $\rho \partial U/\partial t$ remains), by using Poisson's equation. Upon integrating Eq. (4.2) in the whole space and since the matter distribution is spatially bounded, one obtains the energy integral in two successive forms, by using the fact that the field **g** is $O(1/r^2)$ at large $r$ and $\partial U/\partial t$ tends towards 0 at large $r$:

$$(4.3) \qquad E_m + E_g \equiv \int_{matter} w_m \, dV + \int_{space} w_g \, dV = \text{Const.} = E,$$

$$(4.4) \qquad E_m + E_g = \int_{matter} \left( w_m + \frac{\rho U}{2} \right) dV = \int_{matter} \rho \left( \frac{v^2}{2} + \Pi - \frac{U}{2} \right) dV = \text{Const.} = E \ .$$

These equations allow a clear analysis of the energy transfer from matter to gravitation field and, inside the contribution of matter, from potential to kinetic and internal energy. Thus, starting from an 'unbound' state in which both the (*positive*) gravitational energy and the (negative) potential energy $E_p = \int -\rho U \, dV$ are closer to zero than in the final state, a gravitational concentration of matter will increase the gravitational energy $E_g$ and thus decrease the total energy of matter, $E_m$, Eq. (4.3). But since $w_m$ includes the potential energy, the decrease in $E_m$ is only due to the decrease in $E_p$, $\delta E_p$, and the 'pure' (internal plus kinetic) energy of matter is in fact increased by the amount $- \delta E_p /2$, Eq. (4.4). There is also a conserva-tion equation for momentum in NG, namely:

$$(4.5) \qquad \frac{\partial (\rho \mathbf{v})}{\partial t} + \mathbf{div}\left( \rho \mathbf{v} \otimes \mathbf{v} - \boldsymbol{\sigma} \right) + \frac{1}{4\pi G} \mathbf{div}\left( \mathbf{g} \otimes \mathbf{g} - \frac{\mathbf{g}^2}{2} \mathbf{I} \right) = 0,$$

[here $\mathbf{u} \otimes \mathbf{v}$ is the tensor product, thus $(\mathbf{u} \otimes \mathbf{v})^{ij} = u^i v^j$ ; the vector operator **div** means the Euclidean divergence of a second-order space tensor **t**, with $(\mathbf{div}\ \mathbf{t})^i = t^{ij}{}_{,j}$ in any coordinates deduced from Cartesian ones by a linear transformation; and **I** or **δ** is the identity tensor, $I^{ij} = \delta^{ij}$]. This is the mere rewriting, using the Poisson and continuity equations, of Newton's second law for the continuous medium: Poisson's equation implies that the last term in Eq. (4.5) is simply $-\rho \mathbf{g}$. But, contrary to Eqs. (4.3)-(44) in which the energy of the gravitation field does influence the global balance, the integration of Eq. (4.5) in the whole space gives nothing else than the standard conservation of the total momentum of matter. In other words: *in NG, the gravitation field has energy and this is positive, but it has zero total momentum.*

Equations (4.4) and (4.5) are given in CHANDRASEKHAR [8] though Eq. (4.5) is there in a form which is similar to the 'conservation equation' for the mass space-time tensor **T** of GR,

(4.6) $\qquad [\text{div}_{\boldsymbol{\gamma}} \mathbf{T}]^{\alpha} \equiv T^{\alpha\beta}{}_{;\beta} = 0.$

(**iii**) General relativity

With Eq. (4.6) appears the problem of energy and conservation laws in GR. As explained by LANDAU & LIFCHITZ [12], it « does not in general express the conservation law of anything » (§101), i.e. it cannot be considered as a true conservation equation. The reason is that no Gauss theorem applies to the divergence of a second-order tensor in a curved Riemannian space (see also STEPHANI [28]). In several relevant situations, including that of an asymptotically flat metric $\boldsymbol{\gamma}$ (i.e. a space-time that is Galilean at infinity), Eq. (4.6) nevertheless implies integral conservation laws. In order to obtain such laws, one rewrites Eq. (4.6) in the form of a divergence with respect to a flat space-time metric $\boldsymbol{\gamma}^0$, by adding to **T** a so-called 'energy-momentum pseudo-tensor of the gravitation field', **t** (the expression of which is not unique), thus :

(4.7) $\qquad (\text{div}_{\boldsymbol{\gamma}^0} \boldsymbol{\theta})^{\alpha} \equiv \theta^{\alpha\beta}{}_{,\beta} = 0, \qquad \boldsymbol{\theta} \equiv (-\gamma)(\mathbf{T} + \mathbf{t})$

(the factor $-\gamma$ is there for reasons which are bound to the form of the field equations in GR). That the space-time may be equipped with a global flat metric is, of course, a strong topological assumption; in general, the discussion here is valid in the domain $\Omega$ where the coordinate system $(x^{\alpha})$ in Eq. $(4.7)_1$ is regular, and the metric $\boldsymbol{\gamma}^0$ can be *defined* by: $\gamma^0{}_{\alpha\beta} = \eta_{\alpha\beta}$ ($=\varepsilon_{\lambda} \delta_{\alpha\lambda} \delta_{\beta\lambda}$ with $\varepsilon_0 = 1$, $\varepsilon_i = -1$ for $i=1,2,3$) in these coordinates. Anyway, in GR, the choice of the 'background' metric $\boldsymbol{\gamma}^0$ (local or global) cannot be imposed. The first equation remains valid in any coordinates deduced by a linear transformation from Galilean coordinates of the flat metric $\boldsymbol{\gamma}^0$, and the definition of (any) **t** in GR makes **t** a tensor only under linear coordinate transformations. Due to the factor $-\gamma$, the definition of $\boldsymbol{\theta}$ and the whole set (4.7) are hence covariant only for those transformations from Galilean coordinates of the metric $\boldsymbol{\gamma}^0$ that are both linear and 'unimodular', thus essentially for Lorentz transformations of the space-time (or the domain $\Omega$) equipped with the flat metric $\gamma^0$. In particular, the expression of (any) **t** gives zero in locally geodesic cordinates for the curved metric $\gamma$. These peculiarities allow to state that « it is doubtful that [one of such pseudotensors **t**, namely that proposed by Einstein] can in general describe something which could be called energy » (TRAUTMAN [29]). If the $\theta^{\alpha\beta}$ 's decrease sufficiently fast at infinity[6] (this implies in particular that the metric $\gamma$ asymptotically coincides with the flat one $\gamma^0$, thereby ensuring an 'asymptotic uniqueness' to $\gamma^0$, in so far as the metric $\gamma$ of GR may be considered unique), the integration of Eq. (4.7) over the whole 'space' (spatial section $x^0$ = Const. of the space-time) is possible and gives, owing to the Gauss theorem:

(4.8) $$P^\alpha \equiv \frac{1}{c} \int_{space} \theta^{\alpha 0} \, dx^1 \, dx^2 \, dx^3 = \text{Const.}$$

This is sometimes interpreted in GR as the conservation law for energy ($\alpha$=0) and momentum ($\alpha$=1,2,3; see e.g. LANDAU & LIFCHITZ [12]). In contrast to Eqs. (4.3)-(4.4), however, the complex expression of pseudo-tensor **t** in terms of derivatives of the metric makes it difficult to draw definite conclusions from Eq. (4.8), regarding the energy transfer. Moreover, it does not seem completely clear what should be the physically motivated conditions ensuring the sufficient decrease at infinity for $\theta$. Thus if one assumes a 'time-independent far field', one finds that $P^i$=0 and that $P^0$ is the active mass entering the expression of the metric at large distance (cf. STEPHANI [28]). But, strictly speaking, only a constant gravitation field, thus without any motion of massive bodies, will give a time-independent field- even if this is the far field. According to the spirit of GR, a general (time-dependent) gravitation field should radiate gravitational energy, and here the discussion is less clear-cut. One often uses the linear approximation of GR and then, as observed by STEPHANI [28], « if one tries to calculate not the loss of energy but the total energy of the system emitting quadrupoles waves, the corresponding integrals diverge for r→∞ if the system emits continuously (the whole space is filled by radiation). » Thus in relevant situations which exhibit qualitative differences with NG, the first-order expansions must be checked by complex expansions. A more severe difficulty is the contradiction between the principle of general relativity and the necessity of introducing a flat reference metric, i.e. some equivalent of the Galilean class of the inertial frames in NG, in order to define a concept of energy in GR. This difficulty, and some other reasons, led LOGUNOV et al. [14] to propose a 'relativistic theory of gravitation' (RTG), in which the field equations of GR are supplemented by the De Donder-Fock harmonic condition (Eqs. (2.39-40) in LOGUNOV *et al.* [14]). The RTG is written in generally covariant form, hence the harmonic condition (2.40) is there an *additional field equation* instead of a coordinate condition. Yet the RTG does admit a privileged class of coordinates (hence also a class of reference frames), similar to the Galilean class of NG or rather to the Lorentz class of SR. These are, in the RTG, Galilean coordinates of the 'base Minkowski space', i.e. the space-time equipped with a flat metric $\gamma^0$ [$(\gamma^{\mu\nu})$ in the notation of LOGUNOV *et al.*; in the RTG, the metric $\gamma^0$ is assumed global, and as a consequence of the field equations $\gamma^0$ is unique]. In such coordinates, as also in coordinates deduced by a linear transformation, their generally covariant Eq. (2.40) becomes identical to Fock's harmonic *coordinate condition*. Hence, Galilean coordinates for $\gamma^0$ are also harmonic coordinates for the 'effective Riemannian space', i.e. the space-time equipped with the curved metric $\gamma$ [$(g^{\mu\nu})$ in their notations]. A possible objection to the RTG is then, that the arguments leading to the field equations of GR depend strongly on the equal status enjoyed by all reference frames: if one has a privileged class of reference frames (exchanging by Lorentz transformations), one has less physical reasons to express gravity by generally covariant equations.

---

[6] A sufficient condition is: $\theta^{\alpha 0} = o(1/r^3)$, $\theta^{\alpha 0}{}_{,0} = o(1/r^3)$ and $\theta^{\alpha i} x^i = o(1/r)$ at large $r \equiv (x^i x^i)^{1/2}$.

### 4.2 *The energy conservation in the studied theory with Newton law*

(**i**) The energy of a free test particle, and its time evolution in non-static situations

As in NG, the energy of a mass point appears first as a natural conserved quantity in the case of time-independent gravitation potential. We have in general from Eq. (2.1) with $p_e = \rho_e\, c^2$ and $\beta = p_e/p_e^\infty$:

(4.9) $$\mathbf{g} = -\frac{c^2}{\beta}\, \mathrm{grad}_\mathbf{g}\, \beta = \mathrm{grad}_\mathbf{g}\, U, \quad U \equiv -c^2 \log \beta\ .$$

By the Newton law (3.8) with purely gravitational force ($\mathbf{F}_0=0$), we have also, using Eq. (3.5):

(4.10) $$\mathbf{g} = (c^2 - v^2)\frac{D_0}{Ds}\!\left(\frac{\mathbf{dx}}{ds}\right) + \frac{\lambda}{\beta}\, \mathbf{g}^{-1}\!\cdot\!\left(\frac{\partial \mathbf{g}}{\partial t}\cdot \mathbf{v}\right)$$

where, by definition,

(4.11) $$\frac{dt}{ds} \equiv \frac{\gamma_v}{c\beta(t,\mathbf{x})}, \quad \gamma_v \equiv \left(1 - \frac{v^2}{c^2}\right)^{-\frac{1}{2}}, \quad v \equiv \mathbf{g}_t(\mathbf{v},\mathbf{v})^{1/2}, \quad \mathbf{v} \equiv \frac{1}{\beta(t,\mathbf{x})}\frac{d\mathbf{x}}{dt}.$$

To evaluate the rate of work per unit rest mass, $c\mathbf{g}\cdot(d\mathbf{x}/ds)$, with $\mathbf{g}$ from (4.10), we first observe that

(4.12) $$\mathbf{g}_{t_0}\left(\frac{D_0}{Ds}\left(\frac{d\mathbf{x}}{ds}\right), \frac{d\mathbf{x}}{ds}\right) = \frac{1}{2}\frac{d}{ds}\left[\mathbf{g}_{t_0}\left(\frac{d\mathbf{x}}{ds}, \frac{d\mathbf{x}}{ds}\right)\right]$$

and, as in Eq. (3.7), we find that

(4.13) $$\frac{1}{2}\frac{d}{ds}\left[\mathbf{g}_{t_0}\left(\frac{d\mathbf{x}}{ds}, \frac{d\mathbf{x}}{ds}\right)\right] = \frac{1}{2}\frac{d}{ds}\left[\mathbf{g}_t\left(\frac{d\mathbf{x}}{ds}, \frac{d\mathbf{x}}{ds}\right)\right] - \frac{1}{2}\frac{\partial \mathbf{g}_t}{\partial s}\left(\frac{d\mathbf{x}}{ds}, \frac{d\mathbf{x}}{ds}\right).$$

Now, just as in the Lemma in [2] (p. 127), one shows easily with (4.11) that

(4.14) $$\frac{c^2 - v^2}{2}\frac{d}{ds}\left[\mathbf{g}_t\left(\frac{d\mathbf{x}}{ds}, \frac{d\mathbf{x}}{ds}\right)\right] = \frac{d}{ds}\left(c^2 \mathrm{Log}\,\gamma_v\right).$$

Accounting for Eqs. (4.12-14), we thus obtain with (4.10):

$$\mathbf{g}\cdot\frac{d\mathbf{x}}{ds} \equiv \mathbf{g}_{t_0}\left(\mathbf{g}, \frac{d\mathbf{x}}{ds}\right) = \frac{d}{ds}\left(c^2 \mathrm{Log}\,\gamma_v\right) - \frac{c^2 - v^2}{2}\frac{\partial \mathbf{g}}{\partial s}\left(\frac{d\mathbf{x}}{ds}, \frac{d\mathbf{x}}{ds}\right) + \frac{\lambda}{\beta}\frac{\partial \mathbf{g}}{\partial t}\left(\mathbf{v}, \frac{d\mathbf{x}}{ds}\right),$$

or, using (4.11):

(4.15) $$\mathbf{g}\cdot\frac{d\mathbf{x}}{ds} = \frac{d}{ds}\left(c^2 \mathrm{Log}\,\gamma_v\right) + \left(\lambda - \frac{1}{2}\right)\frac{\partial \mathbf{g}}{\partial s}(\mathbf{v},\mathbf{v}).$$

Combining Eqs. (4.9) and (4.15), we get

(4.16) $$\frac{d}{dt}\left(c^2 \mathrm{Log}\,\gamma_v\right) + \left(\lambda - \frac{1}{2}\right)\frac{\partial \mathbf{g}}{\partial t}(\mathbf{v},\mathbf{v}) = \mathbf{g}\cdot\frac{d\mathbf{x}}{dt} = \left(\mathrm{grad}_{\mathbf{g}_{t_0}} U\right)\cdot\frac{d\mathbf{x}}{dt} = U_{,i}\frac{dx^i}{dt} \equiv \frac{dU}{dt} - \frac{\partial U}{\partial t}.$$

In the static case ($\beta_{,0}=0$, whence $U_{,0}=0$ and $\mathbf{g}_{,0}=0$) we have thus, coming back to the expression (4.9)$_2$ of $U$:

$$\text{(4.17)} \qquad \frac{d}{dt}\left(c^2 \text{Log}\, \gamma_v + c^2 \text{Log}\, \beta\right) = 0, \qquad \gamma_v \beta = \text{Const.}$$

But the expression of the (internal plus kinetic) energy in SR is

$$\text{(4.18)} \qquad e_{pm} = m(v)\, c^2 = m(0)\, \gamma_v\, c^2,$$

hence we have got the result that the total energy of the test particle, including its potential energy in the gravitation field, is simply

$$\text{(4.19)} \qquad e_m = e_{pm}\, \beta \; (= \text{Const. if } \beta_{,0} = 0).$$

As in NG, this total energy is lower than the 'pure' energy $e_{pm}$, in other words the gravitational (potential) energy of *matter*, $e_{gm}=e_m-e_{pm}$, is negative. The constancy of $e_m$ in the static case is also a result of GR (cf. LANDAU & LIFCHITZ [12], § 88), since $\beta=(\gamma_{00})^{1/2}$ [Eq. (2.12)]. Yet in GR, this result is not derived from a Newton law and therefore the evolution of $e_m$ in the time-dependent case is complex (moreover, the definition of $e_m$, which involves only $\gamma_{00}$ and is hence non-covariant, does not make sense in the general case in GR, nor in the RTG). Here the evolution of $e_m$ is simple, especially if $\lambda=1/2$; by Eq. (4.16) we get:

$$\text{(4.20)} \qquad \frac{d}{dt}\left(\text{Log}(\gamma_v \beta)\right) = \frac{\partial}{\partial t}(\text{Log}\, \beta) + \frac{1-2\lambda}{2c^2}\frac{\partial \mathbf{g}}{\partial t}(\mathbf{v},\mathbf{v}),$$

$$\text{(4.21)} \quad \frac{d}{dt}(\beta \gamma_v) = \gamma_v \frac{\partial \beta}{\partial t} + \beta \gamma_v \frac{1-2\lambda}{2c^2}\frac{\partial \mathbf{g}}{\partial t}(\mathbf{v},\mathbf{v}), \text{ or } \frac{de_m}{dt} = \frac{e_m}{\beta}\frac{\partial \beta}{\partial t} + \frac{1-2\lambda}{2c^2}e_m \frac{\partial \mathbf{g}}{\partial t}(\mathbf{v},\mathbf{v}).$$

For a light-like particle (photon) one defines $e_{pm}=h\nu$ with $\nu$ the frequency as measured with the local time $t_\mathbf{x}$ of the momentarily coincident clock of the frame E, and one then assumes the Newton law (3.8) with $e_{pm}$ (or rather $e_{pm}/c^2$) in the place of the inertial mass $m(v)$. Defining also $e_m \equiv \beta\, e_{pm}$ for a photon, a similar way of reasoning[7] shows that *exactly the same balance equation (4.21)$_2$ is obtained*. This extends the result of the static case [3], i.e. $e_m$ = Const. Finally, let us come back to the case of an ordinary free test particle, but assuming now the space-time geodesic equation $G^\alpha=0$ of metric $\boldsymbol{\gamma}$ (as in GR). Using, in the special case where an isopotential coordinate system is bound to the privileged frame E, the calculation method of ref. [3], one easily rewrites the 'time part' $G^0=0$ as

$$(4.22) \qquad \frac{d}{dt}(\beta\,\gamma_v) = \gamma_v\,\frac{\partial \beta}{\partial t} - \frac{\beta\,\gamma_v}{2c}\,g_{ij,0}\,v^i v^j\,.$$

This is the same as Eq. (4.21) with $\lambda=1$.

In summary, if one assumes a Newton law (3.8), then the balance equation (4.21)$_2$ is derived for the total energy $e_m$ of any free test particle (mass point or light-like particle), which includes its potential energy in the gravitation field. Unless $\lambda=1/2$ is assumed, a term involving both $e_m$ and the time-variation of the metric is obtained, which will be seen to be incompatible with the obtainment of a local conservation equation, i.e. a balance equation without source term. The value $\lambda=1/2$ is the only one for which Leibniz' rule holds true for the vector time derivative (3.5) entering the Newton law (3.8), but it is incompatible with geodesic motion, which rather corresponds to $\lambda=1$.

(**ii**) The balance equation of energy for dust
We first consider dust, since we have an expression of the energy balance for a free test particle [Eq. (4.21)]. Dust is a continuum made of non-interacting mass particles, each of which conserves its rest mass, so that we may use both Eq. (4.21) and the continuity equation for the density of rest mass, expressing the conservation of the rest mass. The latter means that the rest mass $\delta m_0$ contained within any given 'substantial' domain $\delta\omega$ of the continuum is constant, and is thus a statement that can be expressed in terms of any consistent space and time metric (in so far as the very notion of rest mass is taken for granted). If we follow the motion of the continuum from the privileged frame, we can define its velocity $\mathbf{v}_0$ with the absolute time, and we can use the volume measure $dV^0$ associated with the natural metric $\mathbf{g}^0$ on M, thus

$$\mathbf{v}_0 = d\mathbf{x}/dt, \qquad dV^0 = \sqrt{g^0}\, dx^1\, dx^2\, dx^3.$$

---

[7] Express the Newton law in terms of $D_{1/2}\,(e_{pm}\,\mathbf{v})/Dt_{\mathbf{x}}$, allowing to differentiate $\mathbf{g}_t\,(\mathbf{v},\mathbf{v})\equiv c^2$ using Leibniz' formula with $\mathbf{g}_t$. This gives an expression of $e_{pm}\,\mathbf{g}.\mathbf{v}$. Equate this, as in Eq. (4.16), to that derived from Eq. (4.9).

We then have the usual continuity equation:

(4.23) $$\frac{\partial \rho_{00}}{\partial t} + \mathrm{div}_0(\rho_{00}\mathbf{v}_0) = 0, \qquad \rho_{00} \equiv \delta m_0/\delta V^0, \mathrm{div}_0 \equiv \mathrm{div}_{\mathbf{g}^0},$$

which leads to the following expression for a 'substantial' derivative:

(4.24) $$\frac{d\psi}{dt}\rho_{00} = \frac{\partial}{\partial t}(\psi \rho_{00}) + \mathrm{div}_0(\psi \rho_{00}\mathbf{v}_0), \qquad \frac{d\psi}{dt} \equiv \frac{\partial \psi}{\partial t} + (\mathrm{grad}_0\, \psi)\cdot\mathbf{v}_0 = \frac{\partial \psi}{\partial t} + \psi_{,i}\, v_0^{\,i}$$

(where the scalar product is of course $\mathbf{g}^0$; but it is true that $\mathbf{g}(\mathrm{grad}_\mathbf{g}\,\psi, \mathbf{w}) = \psi_{,i} w^i$ for any metric $\mathbf{g}$ and in any coordinates). We apply this to $\psi = \beta \gamma_v$. According to Eqs. (4.18) and (4.19), $\psi$ is the (rest) mass density of the total energy $e_m$ of dust matter:

$$\delta e_m/c^2 = \beta \gamma_v\, \delta m_0 = \rho_{00}\, \beta \gamma_v\, \delta V^0 \equiv \varepsilon_m\, \delta V^0,$$

hence $\varepsilon_m$ is the volume density of the energy $e_m$ in the frame E, the volume $\delta V^0$ being expressed with 'uncontracted' rods, i.e. in terms of the natural metric $\mathbf{g}^0$ which is not affected by gravity. Combining Eqs. $(4.21)_1$ and (4.24), we get the energy balance for the dust continuum:

(4.25) $$\frac{\partial \varepsilon_m}{\partial t} + \mathrm{div}_0(\varepsilon_m \mathbf{v}_0) = \rho_{00}\, \gamma_v\, \frac{\partial \beta}{\partial t} + \varepsilon_m \frac{1-2\lambda}{2c^2}\, \frac{\partial \mathbf{g}}{\partial t}(\mathbf{v},\mathbf{v}).$$

It is necessary to make contact with the mass tensor **T**, which may be defined for a perfect barotropic fluid as in SR (and as in GR also), thus (cf. FOCK [10]):

(4.26) $$T^\lambda_{\ \mu} = (\mu^* + p/c^2)\, u^\lambda u_\mu - (p/c^2)\, \gamma^\lambda_{\ \mu}, \quad (\gamma^\lambda_{\ \mu} = \delta^\lambda_{\ \mu}),$$

$$\mu^* \equiv \rho^*(1+\Pi/c^2), \quad \rho^* = \rho^*(p), \quad \Pi \equiv \int_0^p \frac{dq}{\rho^*(q)} - \frac{p}{\rho^*(p)},$$

(4.27)  $\quad u^\lambda \equiv dx^\lambda/ds, \quad u^\lambda u_\lambda = 1 \quad$ and, if $x^0 = ct$: $\quad u^0 = \gamma_v/\beta, \quad u^i = (u^0/c)\, dx^i/dt$

and the same for the $T^{\lambda\mu}$ (or $T_{\lambda\mu}$) components, though with $u^\lambda u^\mu$ (or $u_\lambda u_\mu$) instead of $u^\lambda u_\mu$, and with $\gamma^{\lambda\mu}$ (or $\gamma_{\lambda\mu}$) instead of $\gamma^\lambda{}_\mu$, all indices being raised or lowered with metric **γ**. Here, $\rho^*$ is the proper density of rest mass, i.e. the density of rest mass, as measured in a frame C that is bound to the continuum, at least locally and momentarily (comoving frame). In order to find the connection between $\rho^*$ and the density of rest mass $\rho_0$ which would be evaluated by observers in the frame E, one just has to use SR which holds true locally. The local observers in E and in C relate their space and time measurements by local Lorentz transformation, hence the measures, in E, of the 'substantial' volume element in C and the mass density are

(4.28)  $\quad \delta V_E = \delta V_C / \gamma_v, \quad\quad \rho_0 \equiv \dfrac{\delta m_0}{\delta V_E} = \gamma_v \dfrac{\delta m_0}{\delta V_C} = \gamma_v \rho^*$

(note that the observer in E uses the 'true' simultaneity, hence he finds the same relation $\delta V_E = \delta V_C / \gamma_v$ if he simply remembers that the measuring rods of the observer in C are 'truly' Lorentz-contracted in direction **v**). Since the physical metric **g** in the frame E is affected by gravitational contraction of measuring rods, the corresponding volume $\delta V_E$ is dilated with respect to the volume $\delta V^0$ expressed in terms of the natural metric [cf. Eq. (2.7)], thus from (4.23) and (4.28):

(4.29)  $\quad \delta V^0 = \beta\, \delta V_E, \quad\quad \rho_{00} = \rho_0/\beta = \gamma_v \rho^*/\beta.$

Now, from Eqs. (4.26) and (4.27), we have for dust ($p=0$, $\Pi=0$):

$$\rho^* \gamma_v^2 = (T^0{}_0)_E, \quad\quad \rho^* \gamma_v^2 v_0^i/c = (T^i{}_0)_E \quad (x^0=ct),$$

whatever space coordinates bound to E are used; indeed, one has always $\gamma_{0i}=0$ in such coordi-nates [see Eq. (2.8)], hence $u^0 u_0 = \gamma_{00}(u^0)^2 = \gamma_v^2$. By (4.29), we have $\rho^* \gamma_v^2 = \rho_{00}\, \gamma_v\, \beta = \varepsilon_m$, hence we may rewrite the balance equation (4.25) as:

$$(4.30) \quad cT^{\alpha}{}_{0,\alpha} \equiv c\left(\mathrm{div}_{\gamma^0}\mathbf{T}\right)_0 \equiv c\left(T^0{}_{0,0} + \mathrm{div}_0\left(T^i{}_0\,\mathbf{e}_i\right)\right) = \frac{T^0{}_0}{\beta}\frac{\partial\beta}{\partial t} + T^0{}_0 \frac{1-2\lambda}{2c^2}\frac{\partial\mathbf{g}}{\partial t}(\mathbf{v},\mathbf{v})$$

where $x^0=ct$, the space coordinates are bound to E, and the space-time metric $\gamma^0$ is defined by $(ds^0)^2=(dx^0)^2-(dl^0)^2$ with $dl^0$ the line element of the natural space metric $\mathbf{g}^0$; the first identity is valid in any coordinates linearly bound to Galilean coordinates of the metric $\gamma^0$ [if $(M, \mathbf{g}^0)$ is Euclidean. Except for the first identity, Eq. (4.30) is yet valid also if $(M, \mathbf{g}^0)$ is a space with constant curvature].

Equation (4.30) [or (4.25)] still contains the two source terms on the right-hand side. If it transforms into a true conservation equation, this must include the energy of the gravitation field itself. As in NG, we have to use the field equation in order to replace the energy density by a combination of derivatives of the potential [here $f \equiv \beta^2$, with $\beta = p_e/p_e^\infty$, is the exact analogue of the Newtonian potential, Eq. (2.12)]. However, it remains some ambiguity as regards the precise definition of the mass-energy density $\rho$ in the field equation (2.2) or (2.15): is that $T^0{}_0$, $T_{00}$ or $T^{00}$? To answer this question, we will precisely impose the condition that a conservation equation must be obtained, and we will check the interpretation of $\rho$ by the physical consequences of this interpretation. In the case where the model of universe is static at infinity, in the rather wide sense [see after Eq. (2.13)] that $p_e^\infty$ does not depend on the time $t$, we may use the field equation in the form (2.15). We obtain from (2.15) and (2.12), by using (locally) Cartesian coordinates (i.e. $g^0{}_{ij}=\delta_{ij}$ and $g^0{}_{ij,k}=0$ at the considered point):

$$(4.31) \quad \frac{8\pi G}{c^2}\rho\, f_{,0} = \left(f_{,0}f_{,i}\right)_{,i} - \frac{1}{2}\left[\left(\frac{f_{,0}}{f}\right)^2 + f_{,i}f_{,i}\right]_{,0} = -\frac{2}{c^2}\mathrm{div}_0\left(f_{,0}\,\mathbf{g}\right) - \frac{1}{2}\left[\left(\frac{f_{,0}}{f}\right)^2 + \frac{4}{c^4}\mathbf{g}^2\right]_{,0}$$

[where the scalar square $\mathbf{g}^2$ is in terms of the natural metric, $\mathbf{g}^2=\mathbf{g}^0(\mathbf{g},\mathbf{g})$]. This strongly recalls the 'pure gravitational' terms in the Newtonian Eq. (4.2), though with an additional term $(f_{,0}/f)^2$. Now we observe that the first source term on the right of Eq. (4.30) can be written as

$$(4.32) \quad \frac{T^0{}_0}{\beta}\frac{\partial\beta}{\partial t} = \frac{T^0{}_0}{2f}\frac{\partial f}{\partial t} = \frac{T^{00}}{2}\frac{\partial f}{\partial t}.$$

Hence, if we interpret the mass-energy density $\rho$ as $T^{00}$, and if we demand that Leibniz' rule applies to the definition of a vector time derivative, i.e. $\lambda=1/2$ in Eqs. (3.5) and (3.8), then we obtain for dust, from Eqs. (4.30-32), the following *conservation equation for the energy*:

$$(4.33) \qquad \left(\mathrm{div}_{\boldsymbol{\gamma}^0}\mathbf{T}\right)_0 + \frac{1}{8\pi G}\left\{\left[\frac{\mathbf{g}^2}{c^2} + \frac{c^2}{4}\left(\frac{f_{,0}}{f}\right)^2\right]_{,0} + \mathrm{div}_0\left(f_{,0}\,\mathbf{g}\right)\right\} = 0.$$

As will be seen, the interpretation of $\rho$ as $T^{00}$ means that the gravitation field definitely reinforces itself and is thus very plausible. Other possible interpretations: $\rho = T^0{}_0$ or $\rho = T_{00}$, certainly do not allow to rewrite the energy balance (4.30) as a true conservation equation if $\lambda=1/2$ [because one then adds to Eq. (4.33) the term $f_{,0}(T^{00}- \rho)/2$, which is clearly not a 4-divergence]. Finally, if a different value is assumed for $\lambda$, not only Leibniz' rule fails to apply, but the mixed source term [the last term in Eq. (4.30)] makes it more than unlikely that a conservation equation could be obtained for the energy. Thus *we assume $\lambda=1/2$ and $\rho = T^{00}$ from now on*.

(**iii**) Extension of the energy conservation equation to general matter behaviour

For general matter behaviour, the constitutive material particles interact with each other and do not necessarily conserve their rest mass. Still, other conservation laws of microphysics hold true in rather general situations, in particular that for the baryon number which was proposed by CHANDRASEKHAR [8] as a substitute for the mass conservation. However, it is considered here that no conservation law is more fundamental than that for energy. Equation (4.33) applies in the absence of gravitation, i.e. in SR, with **T** the mass tensor of any kind of matter and field {incidentally, matter and non-gravitational fields would be of similar nature, i.e. all these would be microscopic flows in ether, according to the concept of constitutive ether (cf. ROMANI [24])}. In the presence of gravity, Eq. (4.33) has been derived, for dust, from the assumed conservation of the rest mass. We therefore postulate that Eq. (4.33) holds true with **T** the (mixed) mass tensor of any kind of matter and non-gravitational field, in the presence of gravitation. Since the total energy, not the rest mass, is thereby conserved, this postulate contains the possibility that (depending on the constitutive equation) *matter can be created or destroyed by its interaction with a variable gravitation field*.

(**iv**) Some local and global consequences of the conservation equation

The density of pure gravitational energy (or rather of its mass equivalent), with respect to the natural volume measure $dV^0$, appears clearly in Eq. (4.33); it is

$$(4.34) \qquad \varepsilon_g = \frac{1}{8\pi G}\left[\frac{\mathbf{g}^2}{c^2} + \frac{c^2}{4}\left(\frac{f_{,0}}{f}\right)^2\right] = \frac{1}{8\pi G}\left[\frac{\mathbf{g}^2}{c^2} + \frac{1}{4}\left(\frac{1}{\tilde{c}}\frac{\partial \tilde{c}}{\partial t}\right)^2\right],$$

where $\tilde{c} \equiv dl^0/dt = cf$ [cf. Eqs. (2.5) and (2.7)] is the 'absolute' velocity of light (and that of waves of the pressure $p_e$, i.e. gravitation waves) in the direction $\mathbf{g}$. Comparing with the Newtonian gravitational energy $w_g$ [Eq. (4.2)], we see that the studied theory, in which the metric varies with time, says that this variation demands energy. On the other hand, it is recalled that the energy density of matter, $T^0{}_0$ with

$$(4.35) \qquad T^0{}_0 = \varepsilon_m \ (= \rho_{00}\,\gamma_v\,\beta \text{ for dust}),$$

which also is relative to $dV^0$, and which enters the balance equation (4.33), includes its 'potential' energy in the gravitation field [cf. Eqs. (4.18-19)]. Let us examine, in the static case, what follows from assuming that $\rho$ on the r.h.s. of the field equation (2.15) is $T^{00} = T^0{}_0/\beta^2$ (as is imposed if one admits that a conservation equation must exist for energy). Since the density of the 'pure' energy of matter, i.e. without its potential energy, is $\varepsilon_{pm} = T^0{}_0/\beta$ (again with respect to $dV^0$), it means that $\rho = \varepsilon_{pm}/\beta$. Due to the 'Poisson equation' (2.15) with $f_{,0}=0$, Eq. (2.16) determines in the general static case the 'active mass' $m$, giving the expression (2.9) of the space-time metric: at large distance from the massive body (which is unique, since any other body would fall, and which is at rest in the ether), we have approximate spherical symmetry, hence

$$(4.36) \qquad f = 1 - 2\,Gm/(c^2 r) + o(1/r)$$

($r$ being the Euclidean distance from the body), with

$$(4.37) \qquad m = \int_{\text{body}} \rho\, dV^0 = \int_{\text{body}} (\varepsilon_{pm}/\beta)\, dV^0$$

This means that the density $\varepsilon_{pm}$ of the 'pure' (internal plus kinetic) energy is reinforced, as regards its contribution to the 'active gravitational' mass $m$, in the very proportion of the local gravitation field $\beta<1$, i.e. *the gravitation field reinforces itself*, as desired. If, instead of $\rho=T^{00}$, we would assume $\rho=T^0{}_0$ or $\rho=T_{00}$, we would get $\varepsilon_{pm}\beta$ or $\varepsilon_{pm}\beta^3$ in the place of $\varepsilon_{pm}/\beta$ in Eq. (4.37), i.e. the gravitation field would have a *weakening* effect on itself. But the notion of active mass, especially its comparison with the passive gravitational mass [which is also the inertial mass, and still coincides with the pure energy of matter, cf. the Newton law (3.8) and Eqs. (4.18-19)], implies a linearity of the acceleration field **g** with respect to the amount of matter. This amount is best represented by the pure energy of matter, thus its density $\varepsilon_{pm}$ does not coincide with $\rho$. The scalar $f$ depends non-linearly on the distribution of $\varepsilon_{pm}$, even in the static case [Eq. (2.15) with $\rho = \varepsilon_{pm}/\beta$], hence the same is true for the vector **g** [Eq. (2.12)]. One therefore cannot isolate the gravitation force exerted by a subdomain $\Omega$ on a mass point, i.e. the contribution of $\Omega$ to the field **g**. Hence, an actio-reactio principle cannot even be defined for the gravitation force, even in the static case. In summary, the studied theory implies (at least in the static case) that the gravitation field really reinforces itself, and this forbids to define the active gravitational mass for a subdomain of the system. As to the global active mass, it may be defined, at least in the static case, and it is then greater than the sum of the pure energy of matter, for the same reason.

In the more general situation of an isolated, but not necessarily static matter distribution, embedded in *Euclidean* space $M$, the integration of Eq. (4.33) over the whole space and its transformation is possible, under the sufficient condition that $(f_{,0})^2$ is $o(1/r^3)$ and **g** is $O(1/r^2)$ at large $r$ (with $r$ the current Euclidean distance from some point in the group of bodies, and using the fact that $f \approx 1$ at large $r$. In assuming that the behaviour at large $r$ is as in NG, one indeed would expect this decrease of **g** like $1/r^2$; actually, this is not the general case, due to the gravitation waves). Using Gauss' theorem on the sphere $r=R$ and making $R$ tend towards $+\infty$, we obtain then

(4.38) $$W_m + W_g \equiv \int_{\text{matter}} \varepsilon_m \, dV^0 + \int_{\text{space}} \varepsilon_g \, dV^0 = \text{Const.} = W,$$

which is the same as Eq. (4.3), though here $\varepsilon_m$ and $\varepsilon_g$ are given by Eqs. (4.35) and (4.34). Due to the presence of the term $(f_{,0}/f)^2$ in $\varepsilon_g$, the reduction of the second integral to an integral restricted to matter seems possible only in the static case. We have then from (2.12) and (2.15):

(4.39) $$\frac{\mathbf{g}^2}{c^2} = \frac{c^2}{4}\left(\text{grad}_0 f\right)^2 = \frac{c^2}{4}\left(\text{div}_0\left(f \, \text{grad}_0 f\right) - f \, \Delta_0 f\right) = -\frac{1}{4}\left(2\,\text{div}_0\left(f\,\mathbf{g}\right) + 8\pi\, G\rho\, f\right).$$

At large $r$, we have Eq. (4.36) plus $\mathbf{g} \approx - (G m / r^2) \mathbf{e}_r$ (with $\mathbf{e}_r$ the unit radial vector, and where $\approx$ means 'equivalent to'), whence from (4.34) and (4.39) by using Gauss' theorem on the sphere $r=R$ and making $R$ tend towards $+\infty$:

$$W_g \equiv \int_{\text{space}} \varepsilon_g \, dV^0 = (m - \int_{\text{body}} \rho f \, dV^0)/4.$$

Since $\varepsilon_m = \rho f = \varepsilon_{pm} \beta \, (= T^0{}_0)$, we can thus rewrite the conserved energy, in the static situation (which may be the case for the initial, as well as for the final state), as

$$(4.40) \qquad W = \frac{3}{4} \int_{\text{body}} \varepsilon_m \, dV^0 + \frac{1}{4} \int_{\text{body}} \frac{\varepsilon_m}{f} \, dV^0.$$

It is smaller than the active mass $m$ [Eq.(4.37)], and the latter does not need to remain the same in any couple of initial and final states, both being assumed static. Furthermore, and in contrast with NG [see the discussion after Eq. (4.4)], the structure of Eq. (4.40) does not seem to imply that the pure energy of matter, $\int_{\text{matter}} \varepsilon_{pm} \, dV^0$, is necessarily increased as the pure gravitational energy $W_g$ is increased, i.e. in the case of gravitational concentration of matter.

## 5. Conclusion

Although less immediately than in the case of a static gravitation field, a consistent Newton law can still be defined in the general situation within the present theory (and in fact in any theory which, in a reference frame, provides us with a space metric and a time metric). It is consistent in that it is based on a true vector derivative of the momentum, obeying Leibniz' rule with the variable metric $\mathbf{g}$ (here $\mathbf{g}$ is the physical space metric in the privileged frame bound to ether); moreover, this time derivative coincides with the usual absolute derivative in the case of a time-independent metric $\mathbf{g}$, i.e. in the static case. The requirement that Leibniz' rule must hold true permits to select the relevant vector derivative, i.e. $\lambda=1/2$ in Eq. (3.5), from a one-parameter family of candidates. Among the possibilities which are thereby eliminated, the value $\lambda=1$ is likely to correspond to Einstein's motion along space-time geodesics; anyhow, $\lambda=1$ does correspond to geodesic motion in the important case where 'isopotential' coordinates bound to ether still exist (see the definition near Eq. (2.7); this case includes spherical symmetry around a point bound to ether). A general Newton law is important for the status of the theory, since it means that the acceleration field $\mathbf{g}$ [Eq. (2.1)] keeps a direct physical meaning in the most general case. Thus the above result implies a crucial departure of the theory from the 'logic of space-time'. The latter is the commonly accepted *interpretation* of SR, even though the 'logic of absolute motion' can be vindicated as well in SR [4, 6-7, 11, 21-23]. The logic of space-time is, however, *essential* in GR, and even also in its various formulations or modifications as a field theory in flat space-time (including the theory proposed by LOGUNOV *et al.* [14]).

It seems therefore natural to hesitate before taking the new direction at this bifurcation in the studied theory. But another argument leads to the same choice, and this is not quite a detail: if one wants to have a true conservation equation for energy in this theory, one also must assume the Newton law (with Leibniz' rule, thus $\lambda=1/2$) and hence forget space-time geodesics. One then indeed obtains for energy a local balance equation without source term, and this is in terms of a flat space-time metric. The gravitational energy is thereby unambiguously defined (in contrast with GR). Just like in NG, one must recognize that one part of gravitational energy is embedded in matter, as its (negative) potential energy in the gravitation field, while the other (positive) part is present in the whole space and may be called the pure gravitational energy. The expression of the latter in the present theory [Eq. (4.34)] contains just the Newtonian expression, plus a term bound to the fact that any time-dependence of the gravitation field implies a time-dependence of the local space and time standards: according to the studied theory, such variation demands an energy supply. This feature of the theory will be worth to discuss in connection with the question of gravitation waves. When the local balance equation can be integrated (which is not the case if gravitation waves are filling the space), it implies the conservation of the global energy [Eq. (4.38)], which is the sum of the energy of matter (including its negative potential energy) and the pure gravitational energy. The gravitation field does reinforce itself in a direct sense [Eq. (4.37)].

According to the present ether theory, gravitation would have some rather concrete aspects. But it is once again emphasized that the theory is non-covariant, which is a risk as regards its application to celestial mechanics. One thus has to study what would be the effect, say on the motion of a planet considered as a test particle, of a *uniform* motion of the attracting body. Since this effect is at most in $u^2/c^2$ with $u$ the corresponding constant velocity, the perturbation of the Newtonian analysis is likely to be small enough so that the theory, at least, is not *worse* than NG for mechanics of the solar system- except perhaps for time scales beyond the 'horizon of predictability' implied by chaotic behaviour of N-bodies problems. It *might* happen, however, that (due to the effect of the velocity $u$), the theory could fail to account for such very small effects as the advance of perihelion of Mercury, i.e. it could fail to improve NG in that respect (the motion of the perihelion is very sensitive to almost any kind of perturbation). As to the gravitational red shift, the deflection of light rays and the delay of radar signals, it is likely that the effect of the velocity *u* does not change significantly their magnitude, which would thus be correctly predicted. The writer has already verified this point as regards the red shift. Anyhow, one testable difference with GR is already known: no Lense-Thirring effect, i.e. no « gravimagnetic field » due to the self-rotation of a massive body does appear in the present theory [3].

## ACKNOWLEDGEMENTS

I am very grateful to Profs. P. Guélin and E. Soós for discussions which helped me to realize that a consistent energy concept is even more necessary in a non-covariant theory like the present one.

Laboratoire 'Sols, Solides, Structures' [associated with the Centre National de la Recherche Scientifique], Institut de Mécanique de Grenoble, B.P. 53 X, F-38041 Grenoble cedex, France.